\documentstyle[aps,prl,floats,epsf]{revtex}

\def\La40{La$_{1.2}$Sr$_{1.8}$Mn$_{2}$O$_{7}$}
\def\Lax{La$_{2-2x}$Sr$_{1+2x}$Mn$_{2}$O$_{7}$}
\def\Tc{$T_{C}$}
\def\TKT{$T_{\text{KT}}$}
\def\Leff{$L_{\text{eff}}$}

\begin{document}
\draft
\wideabs{

\title{Observation of Kosterlitz-Thouless spin correlations in the colossally
       magnetoresistive layered manganite La$_{1.2}$Sr$_{1.8}$Mn$_{2}$O$_{7}$}

\author{
         S. Rosenkranz,$^1$
         R. Osborn,$^1$
         L. Vasiliu-Doloc,$^{2,3,}$\cite{Lida}
         J. W. Lynn,$^{2,3}$
         S. K. Sinha,$^4$ and J. F. Mitchell$^1$
        }
\address{
          $^1$ Materials Science Division, Argonne National Laboratory,
          Argonne, Illinois 60439
         }
\address{
          $^2$ NIST Center for Neutron Research, National Institute of
          Standards and Technology, Gaithersburg, Maryland 20899
         }
\address{
          $^3$ Department of Physics, University of Maryland,
          College Park, Maryland 20742
         }
\address{
          $^4$ Advanced Photon Source, Argonne National Laboratory,
          Argonne, Illinois 60439
         }

\date{September 3, 1999}

\maketitle

\begin{abstract}
The spin correlations of the bilayer manganite
La$_{1.2}$Sr$_{1.8}$Mn$_{2}$O$_{7}$ have been studied using neutron
scattering. On cooling within the
paramagnetic state, we observe purely two-dimensional behavior with a
crossover to three-dimensional scaling close to the ferromagnetic
transition. Below $T_C$, an effective finite size behavior is observed.
The quantitative agreement of these observations with the
conventional quasi two-dimensional Kosterlitz-Thouless model indicates that
the phase transition is driven by the growth of magnetic correlations, which
are only weakly coupled to polarons above $T_C$.
\end{abstract}

\pacs{PACS numbers: 75.25.+z, 72.15.-v, 75.40.Cx}
} % end \wideabs

% General Intro: CMR, outstanding issues
It is now generally accepted that colossal magnetoresistance (CMR)
in doped manganese oxides involves a strong coupling between spin,
charge, and lattice degrees of freedom \cite{Millis96,Roeder96} close
to the ferromagnetic transition. However, the nature of the transition
and the relative importance of the different interactions in
controlling the magnetotransport above the critical temperature
\Tc\ have not been clearly established.  A number of studies have
indicated that the magnetic phase transition may be unconventional in
CMR compounds, with the observation of an anomalous spin diffusion
component below \Tc\ and a non-divergent correlation length \cite{unconv}.
This has been attributed to the development of magnetic polarons above
\Tc\ and their possible persistence below \Tc, although it can be difficult
to distinguish them from  standard critical fluctuations. In other studies, 
conventional critical scaling of bulk properties has been observed
\cite{conv}. It is important, therefore, to study the nature of the phase 
transition in CMR compounds, and determine if unconventional magnetic 
correlations are essential to the mechanism of CMR.

Naturally layered manganites are derived from the perovskite structure
of the three-dimensional (3D) compounds by the addition of non-magnetic
blocking layers \cite{Moritomo96,Mitchell97}. The bilayer compounds \Lax, 
in which $x$ represents
the hole concentration on the MnO$_2$ planes, have been extensively studied
in recent years because of the insights they provide into the mechanisms
of CMR \cite{layered,Osborn98,Doloc99,Gordon99,Rosenkranz99,Medarde99,%
Fujioka99,Chatterji,Li99}.
The motivation of the present study is to utilize the low-dimensionality
of these compounds to perform a detailed investigation of the spin
correlations close to \Tc\ using neutron scattering. The reduced
dimensionality of the spin fluctuations extends the temperature region
over which critical fluctuations may be studied, and makes them easier
to distinguish from other dynamic processes in the sample \cite{Osborn98}.
This has allowed us to compare the temperature evolution of magnetic
correlations with other two-dimensional (2D) systems exhibiting
similar magnetic ordering.

% previous work
In previous work on the 40$\%$ hole-doped bilayer system we observed strong
2D in-plane ferromagnetic fluctuations above \Tc, with
evidence of competing ferromagnetic and antiferromagnetic interactions 
perpendicular to the planes \cite{Osborn98}. The in-plane correlation length 
$\xi$ was measured in
scans through the 2D rods of magnetic scattering chosen to
optimize the energy integration. However, these measurements are
necessarily performed away from the wavevector corresponding to 3D
magnetic ordering, and so were not sensitive to a possible crossover
to 3D correlations.

% Results
We have now investigated the correlations close to the 3D ordering
wavevector as a function of temperature, both above and below \Tc.
We show that the spin correlations are {\it quantitatively} consistent
with a quasi-2D XY model, exhibiting Kosterlitz-Thouless correlations
above \Tc\ with a crossover to 3D correlations at a correlation length
consistent with the known in-plane and interbilayer exchange constants.
This agreement with other 2D systems indicates that the magnetic
correlations develop conventionally, and that the transition  is a true
second-order phase transition, although there is evidence that it is
smeared by weak inhomogeneity. This is in contrast to a magnetic polaron
model, in which the magnetic correlations are strongly bound to charge
degrees of freedom. Recent x-ray and neutron scattering results provide
evidence of charge localization and the development of charge
correlations above \Tc\ \cite{Doloc99}. Nevertheless, they do
not have a strong influence on the magnetic correlations above \Tc.
Instead, the incipient charge ordering is preempted by the onset of
ferromagnetic ordering, and the charge correlations collapse at \Tc.

% Experimental
The neutron scattering experiments were performed at the NIST Center for
Neutron Research, using the same single crystal of the 40\% hole-doped
bilayer manganite \La40\ (lattice parameters $a = 3.862$ $\AA$ and
$c = 20.032$ $\AA$ at 125 K) that has already been studied in various
other experiments \cite{Osborn98,Doloc99,Gordon99,Rosenkranz99}.
This compound shows a transition to long-range ferromagnetic order
with the Mn spins aligned entirely within the $a$-$b$ plane at
\Tc\ $\approx$ 113 K,
\begin{figure}
 \centering
 \epsfxsize=8.4cm
 \epsfbox{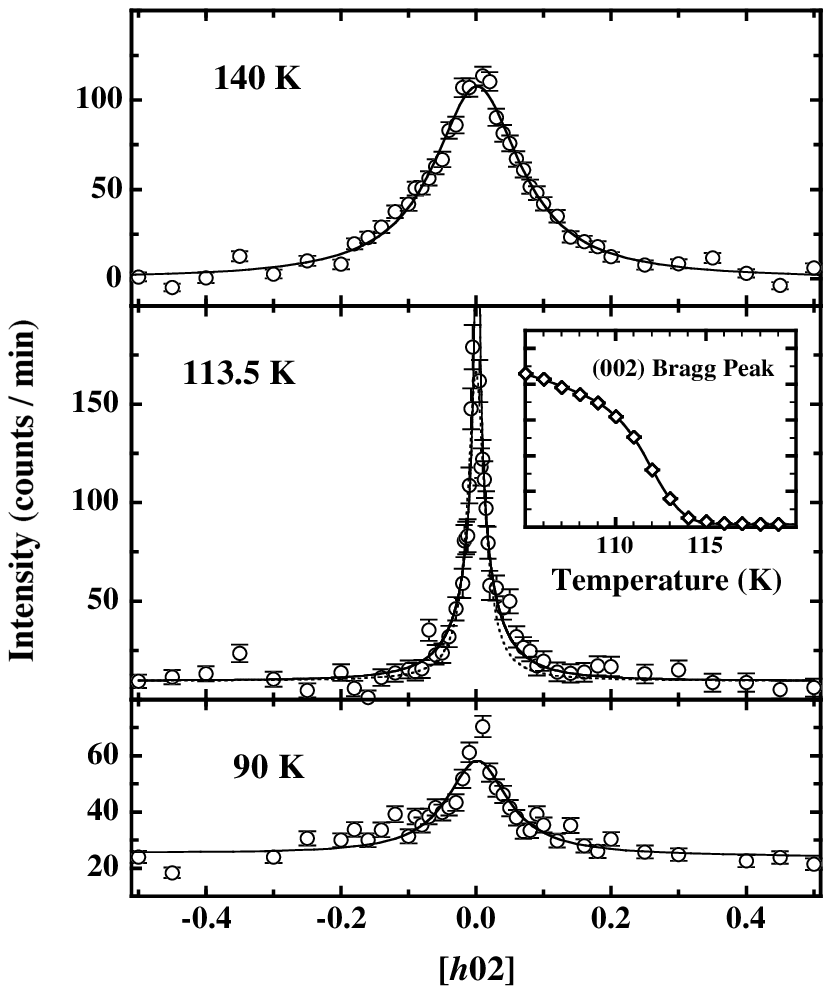}
 \vspace{0.3cm}
 \caption{Magnetic diffuse scattering of \La40 measured along {\bf Q} =
 [$h$02]. The solid lines are the results of the lineshape analysis,
 including a Gaussian \Tc-distribution and the instrumental resolution,
 as explained in the text. The dotted line for $T$ = 113.5 K
 denotes the best least squares fit to a
 Lorentzian profile convoluted with the instrumental resolution
 function. The inset shows the temperature dependence
 of the (002) Bragg peak from Ref.\ \protect\cite{Osborn98} and
 the result of a fit to a standard power-law behavior convoluted
 with a Gaussian \Tc -distribution.
\label{Fig1} }
\end{figure}
coincident with the metal-insulator transition observed in resistivity
measurements. Details on the preparation and characterization of the
sample are given by Mitchell {\it et al.} \cite{Mitchell97}.

The magnetic diffuse scattering was measured on the BT2 triple-axis
spectrometer operating in two-axis mode, i.e., without analyzing
the energy of the scattered neutrons. Scans were taken around the
(002) Bragg position where the nuclear scattering contribution
is extremely weak, with a fixed incident energy 13.7 meV and horizontal
collimations of $60^{\prime}$-$20^{\prime}$-$20^{\prime}$, full width
at half maximum (FWHM). Pyrolitic graphite was used both as monochromator
and filter against higher order contamination. In this two-axis mode,
the diffuse scattering is, in the quasistatic approximation, proportional
to the wavevector-dependent susceptibility $\chi _{T}({\bf q})$, where
${\bf q} = {\bf Q} - \tau$, {\bf Q} is the momentum transfer and $\tau$
denotes a reciprocal lattice vector of the magnetic structure.

% [h,0,2]scans, transition broadening, correlation length

Figure \ref{Fig1} shows the magnetic diffuse scattering observed in scans
along {\bf Q} = [$h$02].
The susceptibility is well described by the usual
Lorentzian profile except for temperatures close to \Tc. In this
region, however, there is also a strong tail in the order parameter
that was attributed to an inhomogeneous broading of the transition
common to disordered systems like the doped manganites \cite{Osborn98}.
Because of the strong dependence of \Tc\ on hole doping
in the bilayer manganites \cite{Medarde99}, such sample inhomogeneities
would manifest themselves as a distribution of \Tc . We have therefore
reanalyzed the scaling behavior of the order parameter, allowing for a
Gaussian distribution of \Tc's. As the inset to Fig.\ \ref{Fig1} shows,
such a distribution is in excellent agreement with the observations.
The newly obtained values of the order parameter are $\beta$ = 0.14(1) and
\Tc\ = 113.2(2) K, compared to $\beta$ = 0.13(1) and \Tc\ = 111.7(2) K
reported in Ref.\ \cite{Osborn98}. The derived standard deviation
$\sigma _{T_{C}}$ = 1.6 K of the Gaussian \Tc -distribution
corresponds to less than 0.4\% variation in the total hole-doping
\cite{Medarde99}. This shows that even a very small sample inhomogeneity
strongly affects the measurements and has to be included for the analysis
of critical quantities.

\begin{figure}
 \centering
 \epsfxsize=8.4cm
 \epsfbox{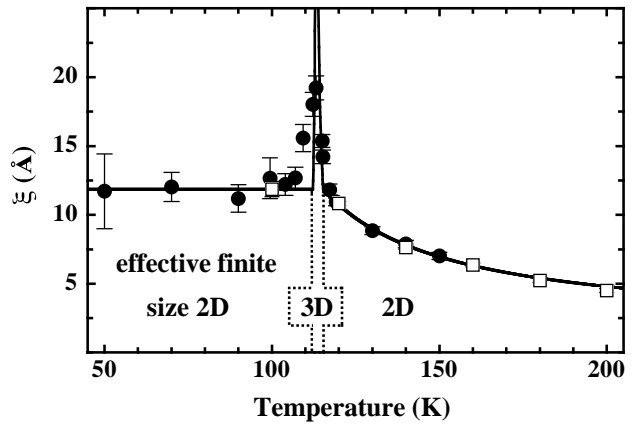}
 \vspace{0.3cm}
 \caption{Temperature dependence of the correlation length $\xi$
  obtained from a Lorentzian profile analysis of the scans
  along {\bf Q}\ = $[h02]$ (circles) and
  {\bf Q}\ = $[1+h,0,1.83]$ (open squares, from
  Ref.\ \protect\cite{Osborn98} ).
  The solid line denotes a fit to the Q2DXY model as explained in
  the text.
\label{Fig2} }
\end{figure}

An analysis of the wavevector dependent susceptibility including the
\Tc-distribution requires a detailed knowledge of the phase transition,
i.e., a model for the temperature dependence of the correlation
length $\xi$ and the static susceptibility $\chi_{T}(0)$, which will be
derived below. For the following discussion, we use the correlation
lengths obtained from a Lorentzian lineshape analysis, which are shown
in Fig.\ \ref{Fig2}.
For these to be valid, the scans need to integrate
over all fluctuation frequencies. In 2D systems, an optimal energy
integration is achieved for a scattering geometry that has the
scattered wavevector parallel to the c-axis \cite{Nielsen76}, but this
is not kinematically possible in scans along {\bf Q} = [$h$02]. However, the
agreement with previous measurements performed in the optimal configuration
at {\bf Q} = [$1+h$,0,1.833] (Ref.\ \cite{Osborn98}) validates the
quasistatic approximation in the present experiments.

% single layer quasi-2D (Q2DXY) model description
In the layered manganites, the separation of the MnO$_2$ bilayers by
insulating (La,Sr)O layers leads to a strong anisotropy, both in transport
properties and magnetic correlations. In a simple nearest-neighbor exchange
Hamiltonian for the magnetic interactions, one therefore expects large
differences between the various exchange constants: $J_{1}$ between spins
within the same plane, $J_{2}$ between spins in different layers within a
bilayer, and $J_{3}$ between spins in different bilayers. This quasi-2D
behavior has been verified by spin-wave measurements
\cite{Rosenkranz99,Fujioka99,Chatterji} which yield
$J_{1}/J_{3} \approx 150$, similar to the anisotropy observed in the
resistivity \cite{Li99}.

To our knowledge, the critical behavior of quasi-2D bilayer systems has not
yet been investigated thoroughly, but there exist detailed theoretical and
experimental studies of single-layer quasi-2D XY magnets (Q2DXY), in which
there is a strong exchange anisotropy $J/J^{\prime}$, where $J$ and
$J^{\prime}$ are the in-plane and interlayer exchange constants, respectively
\cite{Berezinskii73,Hikami80,Bramwell,Nielsen93}. In such systems, both above
and below \Tc, spin fluctuations on length scales that are small compared to
the characteristic length \Leff\ $\approx \surd(J/J^{\prime})$
(in units of the nearest neighbor Mn-Mn distance $a$) are not affected
by the interlayer coupling and are therefore purely 2D. When approaching
\Tc\ from above, one expects the correlation length and the static
susceptibility $\chi_{T}({\bf q}=0)$ to increase according to the
Kosterlitz-Thouless expressions
\begin{equation}
  \label{xiKT}
    \xi/a = \xi_{0} \exp[ b (T/T_{\text{KT}} -1)^{-1/2} ]
\end{equation}
and
\begin{equation}
  \label{chiKT}
    \chi _{T}(0) = C \exp[ B (T/T_{\text{KT}} -1)^{-1/2} ] ,
\end{equation}
where \TKT\ is the topological ordering temperature of the 2D XY model
\cite{KT}. Once the interlayer interaction becomes important, i.e., when
the correlation length reaches the order of \Leff, a crossover to 3D
scaling is expected. Renormalization group theory estimates a 3D ordering
temperature \Tc\, that is related to \TKT\ by \cite{Hikami80}
\begin{equation}
  \label{Tcpred}
    T_{C} = T_{\text{KT}} [ 1 + ( b/\ln L_{\text{eff}})^2 ] .
\end{equation}
Below \Tc, the correlation length decreases rapidly to \Leff, where spin
fluctuations are again unaffected by the interlayer coupling and the
correlation length remains constant. From this point on, although the
magnetization is 3D, the {\it fluctuations} are
2D and the system can be modeled by a 2D system
of effective size \Leff\ \cite{Bramwell}. This behavior has been
observed, for example, in Rb$_2$CrCl$_4$ \cite{Nielsen93}.

% Comparison with Q2DXY ( Fig2, Fig3 )
As is shown in Fig.\ \ref{Fig2}, our observations are in remarkable
agreement with the above predictions of the Q2DXY model. From a least
squares fit of Eq.\ (\ref{xiKT}) to the observed correlation length
above 120 K, we obtain $\xi_{0} = 0.3(1)$, \TKT\ = 64(5) K, and
$b = 2.1(2)$, in excellent agreement with the theoretical value of
$b \approx 1.9$ \cite{Bramwell,Nielsen93}. For the static susceptibility
we obtain $B = 3.9(4)$, very close to the theoretical value
$B = b*(2-\eta)$, where the critical exponent $\eta < 1/4$. Using the
fitted values for $b$ and \TKT\ and $J/J^{\prime} = 150$, we obtain
\Tc\ $\approx$ 109 K, consistent with the value \Tc\ $=113.2$ K derived
from the order parameter.

The deviation from the purely 2D behavior above \Tc\ occurs at a correlation
length $\xi_{\text{2D}}$ $\approx$ 12 $\AA$, which also corresponds to the
effective finite size observed below \Tc. This is in rough agreement
with the value of \Leff\ $\approx$ 40 $\AA$ predicted from the measured
in-plane and interbilayer exchange constants. In the region around \Tc,
we approximate the behavior of both the correlation length and the static
susceptibility with an abrupt crossover to standard 3D scaling, although
a more complete description would have to include a crossover scaling
function \cite{Pfeuty74}.

\begin{figure}
 \centering
 \epsfxsize=8.4cm
 \epsfbox{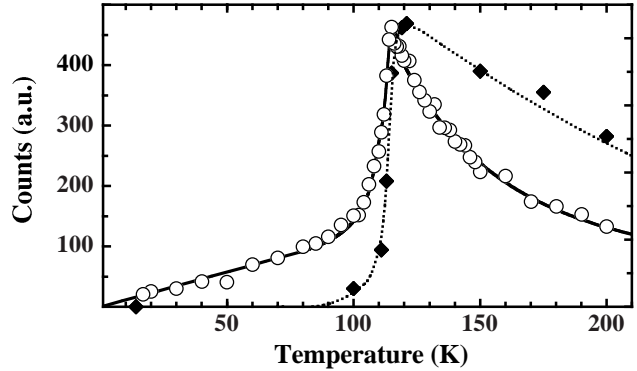}
 \vspace{0.3cm}
 \caption{Temperature dependence of the static susceptibility $\chi_{T}(0)$
  (circles) and quasistatic
  polaronic scattering (diamonds) from Ref. \protect\cite{Doloc99}.
  The solid line denotes a fit to the Q2DXY model as explained in the
  text. The dashed line is a guide to the eye.
 \label{Fig3} }
\end{figure}

% Lineshape analysis including transition broadening
This model for the correlation length and the static susceptibility
(shown as solid lines in Figs.\ \ref{Fig2} and \ref{Fig3} ) can now be used
for an analysis of the peak lineshapes that includes the \Tc -distribution.
Figure \ref{Fig1} shows that this model gives a good description of
the lineshape even
close to \Tc. It is important to note that the model includes a
divergence at \Tc\ of both $\xi$ and $\chi_{T}(0)$, which is not
apparent in the observed data. From a simple Lorentzian
lineshape analysis, one would conclude erroneously that the correlation
length does not diverge at \Tc.

% Discussion
Although the inhomogenous broadening prevents a reliable determination of the
critical scaling within the 3D regime, it is evident from the above analysis
that the magnetic correlations are quantitatively consistent with a 
conventional model of quasi-2D behavior.  To underline the 
significance of this conclusion to our understanding of CMR, we compare
our results with the recent observation of charge correlations in the 
same compound \cite{Doloc99}. In the paramagnetic phase, we have observed
a growth of diffuse x-ray and neutron scattering, arising from the strain
field produced by quasistatic polarons.  Furthermore, broad
incommensurate peaks modulating this diffuse scattering show that
these polarons become increasingly correlated with each other, producing
short range charge ordering on lowering the temperature towards \Tc.
Both the quasistatic polaronic scattering and the charge
correlation peaks start to collapse just above \Tc, and disappear
in the ferromagnetic, metallic state (see Fig.\ \ref{Fig3}).
It has been argued that these polarons explain the low hole mobility in the
paramagnetic state, which cannot be due to double exchange
alone \cite{Millis96}.

The present observations are strong evidence that the phase transition in 
\La40\ is driven by the growth of magnetic correlations, which are only
weakly  coupled to the polarons above \Tc.  The polarons are likely to
induce some exchange disorder, but it is evidently not sufficient to
disrupt the development of critical magnetic fluctuations.  This
conclusion is not consistent with a magnetic polaron model, in which the
spin correlations are strongly coupled to the charge degrees of freedom.
In such models, the ferromagnetic phase transition is induced by a
transition from small to large polarons \cite{Roeder96}, but it is
unlikely that such a transition would mimic the scaling behavior
observed here.  This does not mean that such models cannot be relevant
to other CMR compounds, particularly those with much stronger
electron-phonon coupling, but they are not essential to a description
of the CMR process.

Although the spin-charge coupling is weak above \Tc, the magnetic
correlations ultimately drive the metal-insulator transition; once the
spin correlations extend over a large enough region, the double exchange
interaction can overcome the mechanism responsible for localizing the
charges, inducing the polaron collapse \cite{Doloc99}.  This may 
affect the critical scaling in the 3D regime and could be responsible for the 
unusually low value of $\beta$ (in quasi-2D XY systems, $\beta$ is predicted
to be approximately 0.23 \cite{Bramwell}).

% Conclusion
Our results demonstrate that the critical properties of the bilayer CMR
manganite \La40, both above and below \Tc, are in quantitative
agreement with an effective finite-size 2D XY model. The magnetic
and charge degrees of freedom are therefore only weakly coupled
except close to \Tc, where the growth of magnetic order delocalizes
the polaronic charges. It is possible that some form of charge ordering
would occur at lower temperature if it were not preempted by the
magnetic ordering.
From these observations, we conclude that magnetic polaron models are not 
appropriate to the present bilayer compound and are therefore not
universal to CMR.

%\acknowledgments
This work was supported by the U.S.\ DOE BES DMS W-31-109-ENG-38
and NSF DMR 97-01339.

% references.

\end{document}